\documentclass[prb,aps,twocolumn,showpacs]{revtex4}
\usepackage{amssymb,graphics}

\newcommand{\beq}{\begin{equation}}
\newcommand{\eeq}{\end{equation}}
\newcommand{\bea}{\begin{eqnarray}}
\newcommand{\eea}{\end{eqnarray}}

\begin{document}

\title{
Surface charge relaxation and the pearling 
instability of charged surfactant tubes}
\author{T. T. Nguyen, A. Gopal, K. Y. C. Lee and T. A. Witten}
\affiliation{The James Frank Institute, The University of Chicago,
5640 South Ellis Avenue, Chicago, Illinois 60637}
\begin{abstract}
The pearling instability of bilayer surfactant tubes was recently 
observed during the collapse of fluid monolayers of 
binary mixtures of DMPC$-$POPG and DPPC$-$POPG
surfactants. We suggested it has
the same physics as the well-known Raleigh instability
under the action of the bilayer surface tension whose magnitude
is dictated by the electrostatic interaction between
charged surfactants.
In this paper, we calculate the relaxation of charge molecules
during the deformation of the tubes into pearling structure.
We find the functional 
dependence of the relaxation energy on the 
screening length $\kappa^{-1}$ explicitly.
Relaxation effect lowers the cost of bending a tube into 
pearls making the cylindrical tube even more unstable.
It is known that for weak screening case where the tube 
radius is smaller than the screening length of the
solution, this relaxation effect is important.
However, for the case of strong screening it is negligible. 
For the experiments mentioned, the situation is marginal. 
In this case, we show this relaxation effect
remains small. It gives less than 20\% contribution to the
total electrostatic energy.
\end{abstract}

\maketitle

\section{Introduction}

The formation of surfactant tubes and budding of 
spheroidal structures are of significant interest in 
biological processes. In particular, such structures 
constitute intermediates that are responsible for critical 
cellular processes like material trafficking from the Golgi 
complex\cite{Golgi}, and fusion 
and fission of membranes\cite{translocation}.
As seen during cell locomotion and the 
formation of Golgi structures, natural surfactant tubes are 
prone to transform to a structure resembling a string of pearls.

Pearling has been induced in tubular phospholipid membranes by 
adsorption of oil\cite{OilPearl} or polymer \cite{PolymerPearl},
on the one side of the membranes.
These phenomenon were interpreted in terms of the creation of
membrane spontaneous curvature due to those
external stimulus. 

\begin{figure}[h]
\resizebox{8cm}{!}{\includegraphics{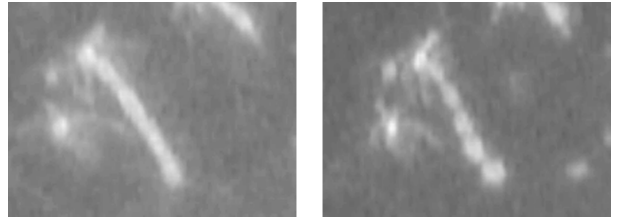}}
\caption{Two snapshots taken within a few seconds
of each other, showing a surfactant tube
undergoing a pearling instability. The monolayer
are 7DMPC:3POPG binary mixture.}
\label{fig:pearlpic}
\end{figure}

We have recently observed pearling
in tubular structures formed 
during the collapse (2D-3D transition) of fluid monolayers
of mixed phospholipids\cite{Ajay}. Collapse 
in binary monolayers of 7DPPC:3POPG and 7DMPC:3POPG lead to 
the formation of cylindrical tubes\cite{ribbon}.
These tubes can be 10s 
of microns in length, with diameters close to 1$\mu$m 
(limit of resolution). A few of these are wide enough to 
resolve detailed features. As seen in Fig. \ref{fig:pearlpic},
such tubes 
show instability towards pearling without the introduction 
of any external gradients that may affect 
or induce the spontaneous curvature. 
Furthermore, the tubes,
being microscopic and submerged in water are likely
to be composed of surfactant bilayers, 
which are in the liquid phase at the temperature measured. 
This suggests that the tube surface does
not have intrinsic spontaneous curvature itself.
Thus the above
mentioned mechanisms of pearling instability is questionable
for the present case.


%

In the same paper, we proposed that the pearling instability
observed is due to
a simpler mechanism. Namely, the instability is
caused by the surface tension energy of the surfactant bilayer.
This is very similar to the well-known Raleigh instability 
of cylinder of fluid. 
Indeed, because one type of surfactants
used in experiments 
are POPG surfactant, a charged molecule, 
the surface tension
of the bilayer should be, at least, of the same order of
magnitude as the electrostatic energy per unit area
of the tube.
The latter is $\pi\sigma_0^2/\kappa D$ where $\sigma_0$ is the
surface charge density of the bilayer, $\kappa$ is the
inverse screening radius of the solution and $D=80$ is
the dielectric constant of water.
Using relevant experimental parameters, 
this electrostatic energy is estimated to be about
$10^{-3}$mN/m.  
On the other hand
the bending energy of a surfactant phospholipid
bilayer is known\cite{BendingRigidity}
to be about $\gamma\simeq 30k_BT$ (here,
$k_B$ is the Boltzmann constant and $T$ is the
temperature of the solution). 
For a tube with radius $R_0$ of 1 micron, 
this translates into an elastic 
(bending) energy of about $\gamma/R_0^2\simeq 10^{-4}$mN/m
per unit area. 
Thus the elastic energy is
more or less negligible in comparison to surface tension energy.
In other words, an instability similar to Raleigh instability
of a fluid cylinder must be present for micron size tubes.
For the case of a fluid cylinder, this instability
leads to the breaking up of the cylinder into small droplets. However, 
for a surfactant tube, the breaking process is
improbable because all surface tension energies involved are
far below
the rupture tensile stress (about 1mN/m) of the lipid bilayer.
The pearled structure is obviously the most likely
candidate for the final structure of this instability.

We
note that
electrostatic surface tension 
energy is a negative quantity (repulsions
between charged molecules prefer to expand, not to reduce, the
surface area of the tube). Therefore, it cannot be the driving
force of instability. However, because the total
surface tension of the bilayer must be positive,
the presence of the negative surface tension forces the 
non-electrostatic surface tension to be positive and greater
than the electrostatic one in magnitude. 
Quantitative calculation of the non-electrostatic surface tension
from the balance between these energies requires the knowledge
of the lateral compressibility of the bilayer and
is presented in details in Ref. \onlinecite{Ajay}. 
It is shown that
this non-electrostatic surface tension is of the same
order of magnitude as the the electrostatic one, and thus,
is much bigger than the bending energy.
It is this
non-electrostatic surface tension whose magnitude is
dictated by the electrostatic counterpart which drives the tube
towards pearling instability.

This explanation of pearling instability as a Raleigh 
instability due to electrostatic effect is
supported further by the experimental
observation that adding monovalent salt
to the solution diminishes electrostatic energy and remove
the instability of the cylinder. 
It is also observed that
cylinders with smaller radius are more stable against pearling.
This can be explained simply by the rapid increase of
the elastic energy when the tube radius decreases.

There is another well known electrostatics-induced
pearling instability in literature, namely, the pearls-on-string
structure of polyelectrolyte in poor solvent or of
polyampholytes \cite{Rubinstein}.
The physics behind these instabilities is different
from that described above.
Even though both instabilities are induced by electrostatic effects,
in surface tension induced pearling,
the characteristic size of pearls is determined kinetically.
On the other hand, the pearl size and period of polyelectrolytes is
determined thermodynamically by the balance between electrostatics
energy and non-electrostatic energy (entropic
or solvent-monomers interactions).
This leads to the strong redistribution
of the charge molecules (strong charge relaxation) in the
system to lower its overall free energy,
which is obviously not needed in the case of
Raleigh instability. One expects charge relaxation is
the driving force of pearling instability when screening of
the solution is weak
such that the Debye screening length is much larger than
the pearl size. On the other hand when the
screening is strong, the Debye screening length
is much smaller than the pearl size, electrostatics is a
short range interaction and the instability is of
the dynamical Raleigh type.

In the experimental system of Ref. \onlinecite{Ajay}, pure water is
used. In this case, the screening radius of the system is
comparable to the tube radius (about one micron). Thus, the situation
is marginal, and it is not clear whether or not the relaxation of charge
surfactants still plays a significant role in the
pearling instability. In this paper, we would like to address
this question by calculating explicitly the gain in the
electrostatic energy of the system when the charged molecules
of the bilayer redistribute themselves during the pearling transition.
We show that in this marginal case where the
screening radius is equal to or smaller than the tube radius, the
charge relaxations remains small. The energy gain
due to this effect contributes at most 20\% to the total electrostatic
energy. For smaller screening radius, the ratio between these
two energies decreases very fast (as fourth power in the
ratio between the screening length and the pearl size). Thus
in the experiments of Ref. \onlinecite{Ajay}, one can
neglect the modulation in surface charge density when considering
the electrostatics of the system.

This paper is organized as follows.
In Sec. II, using linear analysis,
we briefly calculate changes in the surface tension
and bending energies
when a cylinder deforms into a string-of-pearls.
In Sec. III, we calculate the change in the electrostatic
energy under this deformation and separate the
contribution due to the relaxation of charged
molecules. The latter is always negative.
This gives an additional gain in the energy of deformation,
making the tube even more unstable.
In section IV, we discuss the relative 
importance of charge 
relaxation effect as well as various approximations
involved.

\section{Elastic energy}

Even though the elastic (surface tension and bending)
energy changes when a cylinder
undergoes pearling deformation has been 
calculated
\cite{PincusCoiling,Helfrich} before, we briefly
repeat the calculation here in order to 
introduce the notations and to simplify 
their comparison with electrostatic energy
in later sections.

Let us start with a model elastic 
free energy describing the cylindrical tubes.
Denoting the bilayer tube length $L$, area $S$,
and volume $V$, our starting free
energy is the sum of 
the surface tension energy, the bending energy
and an osmotic pressure energy:
\bea
E &=& E_{s}+E_{b}+E_{o}=\alpha \int dS+\nonumber\\
  &&~~~+ \int dS [2 \gamma H^2+\bar{\gamma}K]
  	+\delta p\int dV
\label{totalE0}
\eea
where $\gamma$ and $\bar{\gamma}$
are the bending rigidity and the 
Gaussian bending rigidity
of the bilayer, $H$ and $K$ are the 
mean and Gaussian curvature of the tube surface, $\delta p$ is 
an osmotic pressure difference between the inner
and outer region of the tube. In the above model,
the osmotic pressure term
is somewhat artificial. This
term is needed to make the cylindrical shape the minimum of
the energy for certain range of the parameters
$\alpha$, $\gamma$ and $\delta p$
(because the growth of the tubes is
slow in experiments, we consider the tubes are
in (quasi-)equilibrium and their shapes are determined 
by the minimum of the free energy). Without this term,
a spherical vesicle will always be the 
shape which minimizes the free energy, Eq. (\ref{totalE0}).
This osmotic pressure term was also used by the authors of Ref. 
\onlinecite{Helfrich} to study instability of 
cylindrical vesicles.
A second choice for the model energy 
is to replace the osmotic pressure
term in Eq. (\ref{totalE0}) by a line tension term, which
has been used by the authors of Ref. \onlinecite{PincusCoiling}
to study coiling instability in multilamellar tubes.
Each of these models incorporate different physics in
stabilizing the cylindrical tube. The choice of one
model over the other is not important in this paper because
we do not allow either the volume or the length of the cylinder
to change in our subsequent analysis of the cylinder instability.
In the free energy, Eq. (\ref{totalE0}), the surface tension
and the osmotic pressure are actually Lagrangian multipliers
which enforce the restrictions of area and volume conservation
of the surfactant tube.

\begin{figure}[h]
\resizebox{8cm}{!}{\includegraphics{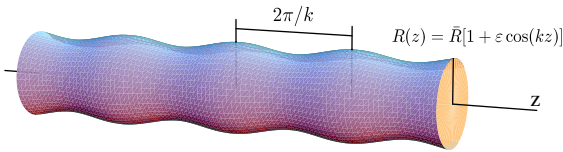}}
\caption{A cylindrical tube is deformed radially with
wavevector $k$.}
\label{fig:DeformedTube}
\end{figure}

Within linear analysis, 
to investigate the change in the energy
of a tube undergoing a pearling instability,
let us slightly deform the cylinder radially
with a relative amplitude, $\varepsilon\ll 1$,
and a wave vector $k$ (see Fig. \ref{fig:DeformedTube}).
As a result, the radius of the new tube varies 
along its axis according to:
\beq
R(z)=\bar{R}[1+\varepsilon \cos(kz)]
\label{Rz}
\eeq

Due to the small permeability of water through the
bilayer surface, in our model, we require that
the tube volume does not change under deformation.
This makes the average radius, $\bar{R}$, of the new tube 
different from the original radius $R_0$. This new radius can
be easily calculated. The volume per unit length of the new tube
is:
\beq
V = \frac{k}{2\pi}\int_{-\pi/k}^{\pi/k} dz
\int_0^{R(z)} 2\pi rdr = \pi \bar{R}^2 (1+\varepsilon^2/2)
\eeq
The condition of volume conservation then leads to the simple
relation:
\beq
\bar{R} = R_0/\sqrt{1+\varepsilon^2/2}
\simeq R_0(1-\varepsilon^2/4)
\label{RR0}
\eeq

Let us calculate the change in the surface
tension energy of the tube.
The element of the area of the new tube is,
\bea
&&dS(z,\phi) =
	dz d\phi~ R(z)\sqrt{1+R^{\prime}(z)^2}
	\nonumber\\
&&~~~=dzd\phi~\bar{R}[1+\varepsilon\cos(kz)]
	\sqrt{1+[\varepsilon k\bar{R}\sin(kz)]^2}
\label{dA}
\eea
where $R^{\prime}(z)\equiv dR/dz$.
The area of the deformed tube per unit length is then,
\bea
A &=&\frac{k}{2\pi}
\int_{-\pi/k}^{\pi/k}\int_0^{2\pi}dS(z,\phi)
\nonumber\\
&=&2\pi R_0\frac{
	E\left[
		-{(kR_0)^2\varepsilon^2}/{(1+\varepsilon^2/2)}
	\right]
	}{(\pi/2)\sqrt{1+\varepsilon^2/2}}
\nonumber\\
&\simeq&
	2\pi R_0 \left[1-\frac{1-(kR_0)^2}{4}\varepsilon^2\right]
~~~{\rm for}~~~\varepsilon \ll 1.
\label{newA}
\eea
where $E$ is the complete elliptic integral of the second kind. 
The change in surface tension energy per unit
length is easily calculated to be:
\bea
\Delta E_{s}&=&\alpha(A-2\pi R_0)
\nonumber\\
&\simeq&
\frac{\pi\alpha R_0}{2}[(kR_0)^2-1]
\varepsilon^2
\label{Esurf}
\eea

Let us next calculate the change in the bending energy. 
Because the Gaussian curvature energy 
$\bar{\gamma}\int dA K$ is a topological invariant quantity and
we do not change the topology of the tube, this energy
does not change. For the
mean curvature, standard geometry consideration\cite{SafranBook}
of the tube surface gives:
\beq
H=\frac{
1+R^{\prime}(z)^2
-R(z)R^{\prime\prime}(z)
}
	{2R(z)[1+R^{\prime}(z)^2]^{3/2}}.
\label{eq:Hz}
\eeq
Substituting Eq. (\ref{eq:Hz}) into the expression for
bending energy (the second term in Eq. (\ref{totalE0})),
and keeping terms up to
second order in $\varepsilon$, one obtains for 
the bending energy change:
\beq
\Delta E_b\simeq 
	\varepsilon^2 \frac{\pi\gamma}{4R_0}
	\left[3+2(kR_0)^4-(kR_0)^2\right].
\label{Ebend}
\eeq
%

%

\section{Electrostatic energy and relaxation of charged
surfactants}

Let us now proceed to calculate the electrostatic energy
change in the system under pearling deformation. 
We use the standard
Debye-H\"{u}ckel (DH) approximation to describe interactions between
the charged surfactant molecules. In this approximation,
the only role of free ions in solution is to
screen the Coulomb potential of a charged
surfactant. In other words, the electrostatic potential due to
a charged surfactant molecule at a distance $r$ from it is
\beq
V^{DH}(r)=e \exp(-\kappa r)/D r,
\label{VDH}
\eeq
where $e$ is the charge of one surfactant molecule (without loss
of generality, we assume the charge of the surfactant is
positive, $D=80$ is the dielectric constant of water),
and
$\kappa$ is the inverse Debye-H\"{u}ckel
screening radius. If the concentration
of monovalent ions is water solution is $c_0$, 
$\kappa$ is given by:
\beq
\kappa = \sqrt{8\pi c_0 e^2/Dk_BT}
\eeq

The deformation of the tube also leads to the redistribution
of charged molecules (or charge relaxation).
The degree of charge relaxation depends on the specific system. 
In this section, for simplicity
we assume the relaxation of 
charged surfactant happens instantly and the distribution
of surface charge is the equilibrium
distribution with respect to a given shape of the tube.
We return to this assumption in detail in
the next section.

To find the equilibrium charge distribution, $\sigma(z)$,
which, in turn,
enables us to calculate the change in the electrostatic
energy, one needs to solve the DH equation
for the electrostatic potential, $V({\mathbf r})$,
in the system
\beq
\nabla ^2 V({\mathbf r}) = \kappa^2 V({\mathbf r})
\eeq
self-consistently with the boundary condition that the electric
field at the tube surface is $2\pi\sigma(z)$ and the
surface charge is at a constant potential. 
In this paper, we use a simpler approach. Namely,
we use a variational approach to calculate the 
electrostatic energy. 
Assuming the following ansatz for the charge
distribution of the deformed tube:
\beq
\sigma(z)=\bar{\sigma}[1+x\cos(kz)],
\label{sigmaz}
\eeq
we optimize the electrostatic energy of the tube 
with respect to the variational parameter $x$.
As we shall see later, $x$ is proportional to $\varepsilon$.
This, coupled with the fact that
for small deformation the response of the system
is linear, the charge distribution obtained using variational
approach is actually the true charge density of the
system up to the second order in $\varepsilon$.

The conservation of the total charge of the tube:
\beq
\sigma_0 2\pi R_0=\frac{k}{2\pi}
	\int_{-\pi/k}^{\pi/k}\int_0^{2\pi}
	dS(z,\phi)\sigma(z)~,
\eeq
immediately gives 
for the average charge density
$\bar{\sigma}$:
\beq
\bar{\sigma}\simeq\sigma_0\left\{
	1+\frac{1-(kR_0)^2}{4}\varepsilon^2-\frac{x\varepsilon}{2}\right\}
\label{sigmabar}
\eeq

The electrostatic energy of the tube with surface
charge density, Eq. (\ref{sigmaz}), is:
\bea
E_e&=&\frac{1}{2}\int dS(z_1,\phi_1) dS(z_2,\phi_2) 
	\sigma(z_1)\sigma(z_2)\times
	\nonumber\\
	&&~~~~~~~~~~V^{DH}[d(z_1,z_2,\phi_1,\phi_2)]
\label{Eelec0}
\eea
where the distance $d(z_1,z_2,\phi_1,\phi_2)$ between the two points
$(z_1,\phi_1)$ and $(z_2,\phi_2)$ on the tube surface is
\bea
&&d^2(z_1,z_2,\phi_1,\phi_2)=
	(z_1-z_2)^2+
	R^2(z_1)+
	\nonumber\\
&&~~~~~
R^2(z_2)-
	2R(z_1)R(z_2)\cos(\phi_1-\phi_2)
\label{dist}
\eea

Substituting Eq. (\ref{dA},\ref{VDH},\ref{sigmaz},\ref{sigmabar}) 
and (\ref{dist}) into
Eq. (\ref{Eelec0}), and expanding the integrand to second
order in $\varepsilon$ ($x$ and $\varepsilon$ are
of the same order of smallness), 
one obtains
the following expression for the electrostatic energy 
per unit length of the tube after integration:
\beq
E_e\simeq\frac{2\pi\sigma_0^2R_0^2}{D}(a_0+\varepsilon^2 a+
	\varepsilon b x + c x^2),
\label{Eelec1}
\eeq
where the coefficients $a_0$, $a$, $b$, and $c$ are:
\bea
a_0 &=& 2\pi I_0(\kappa R_0) K_0(\kappa R_0),\nonumber\\
c&=& \pi I_0\left(\sqrt{\kappa^2+k^2} R_0\right)
	K_0\left(\sqrt{\kappa^2+k^2} R_0\right),
	\nonumber\\
a&=&c+\left(\frac{R_0^2}{4}\frac{\partial^2}{\partial R_0^2}+
	\frac{3R_0}{4}\frac{\partial}{\partial R_0}\right)
	\left(\frac{a_0}{2}+c\right)+
\nonumber\\ && 
	+\frac{\sqrt{\pi}\kappa R_0}{4}
	G^{21}_{13}\left(\kappa^2R_0^2\left|
		\begin{array}{l}
			1 \\
			\frac{1}{2},
			\frac{1}{2},
			\frac{-1}{2}
		\end{array}
		\right.
	\right)
\nonumber\\ && 
	-\frac{\sqrt{\pi(k^2+\kappa^2)} R_0}{4}
	G^{21}_{13}\left[(k^2+\kappa^2)R_0^2\left|
		\begin{array}{l}
			1 \\
			\frac{1}{2},
			\frac{1}{2},
			\frac{-1}{2}
		\end{array}
		\right.
	\right]
\nonumber\\ && 
	+\frac{\sqrt{\pi}\kappa R_0}{2}
	G^{21}_{13}\left(\kappa^2R_0^2\left|
		\begin{array}{l}
			0 \\
			\frac{-1}{2},
			\frac{1}{2},
			\frac{-1}{2}
		\end{array}
		\right.
	\right)
\nonumber\\ && 
	-\frac{\sqrt{\pi(k^2+\kappa^2)} R_0}{2}
	G^{21}_{13}\left[(k^2+\kappa^2)R_0^2\left|
		\begin{array}{l}
			0 \\
			\frac{-1}{2},
			\frac{1}{2},
			\frac{-1}{2}
		\end{array}
		\right.
	\right]
\nonumber\\ 
b&=&2c+R_0\frac{\partial}{\partial R_0}
	\left(\frac{a_0}{2}+c\right),
\label{abc}
\eea
$I_0$ and $K_0$ are the modified Bessel functions of zeroth
order and $G^{mn}_{pq}\left(x\left|
	\begin{array}{l}a_r\\b_s\end{array}\right.\right)$
is the Meijer's G function\cite{Gradsteyn}.

Minimizing the electrostatic energy, Eq. (\ref{Eelec1}),
with respect to $x$, one gets for $x$ and the electrostatic
energy change per unit length:
\bea
x&=&-\varepsilon b/2c.
\nonumber\\
\Delta E_e &=&E_e -\frac{2\pi\sigma_0^2R_0^2}{D}a_0
	=\varepsilon^2 \frac{2\pi\sigma_0^2R_0^2}{D}
	\left[a-\frac{b^2}{4c}\right]
\label{x}
\eea
As expected, $x$ is of the same order of smallness as $\varepsilon$.
This is consistent with the starting assumption
we use in the expansion, Eq. (\ref{Eelec1}).

If the charged surfactant molecules do not relax
to equilibrium surface distribution, 
their density remains constant
under the deformation, $x=0$. From
Eq. (\ref{Eelec1}), the change in the
electrostatic energy in this case is given by:
\beq
\Delta E_e^{\rm norel} = \varepsilon^2
	\frac{2\pi\sigma_0^2R_0^2}{D}a
\label{EelecNorelax}
\eeq
Correspondingly, the energy change due to the 
relaxation of charged surfactants comes from the 
two $x$-dependent terms in Eq. (\ref{Eelec1}):
\beq
\Delta E_e^{\rm rel}=-
\varepsilon^2 \frac{2\pi\sigma_0^2R_0^2}{D}
        \frac{b^2}{4c}
\label{EelecRelax}
\eeq
As one sees from Eq. (\ref{abc}), $c$ is a positive
coefficient. Thus the relaxation energy is negative as
expected: electrostatic
relaxations lower the cost of deforming a tube into pearls.

\section{Discussion}

In this section, 
we comment on the relative importance of various
energies in the system starting with the
electrostatics energy and the contribution
coming from the relaxation of charge molecules
given by Eq. (\ref{x}) and Eq. (\ref{EelecRelax}).
To gain a better physical insight into these equations,
it is instructive to consider the strong screening
case, $\kappa \gg k$ and $\kappa R_0 \gg 1$
and expand the energies in powers of $1/\kappa R_0$.
For $\Delta E_e^{\rm norel}$, the zeroth order 
term of the expansion is
\beq
\Delta E_e^{\rm norel\ (0)}
	\simeq
	        -\frac{\pi^2\sigma_0^2\kappa^{-1}R_0}{2D}
		\left[(kR_0)^2-1\right]\varepsilon^2,
\label{EelecNoRelax}
\eeq
It is easy to see that this energy behaves in the same 
way as the surface tension
energy, Eq. (\ref{Esurf}). One, therefore, identifies
the electrostatic contribution to the surface tension 
of the surfactant tube:
\beq
\alpha_{\rm e}=-\pi\sigma_0^2\kappa^{-1}/D.
\eeq
It is not surprising to see that,
in the absolute value, this ``electrostatic" surface tension
is simply the electrostatic energy per unit area of a
flat bilayer at the same charge density. 
The negative sign in this expression reflects the fact
that electrostatic repulsions between charged surfactants
prefer to increase the area of the surfactant bilayer.
The total surface tension of the layer, of course, remains
positive because of the non-electrostatic interaction
between surfactant molecules counter-balance this
negative electrostatic contribution. 

The next non-zero term of the expansion of $\Delta E_e^{\rm norel}$
is of second order in $1/\kappa R_0$:
\bea
&&\Delta E_e^{\rm norel\ (1)}
        \simeq
	\varepsilon^2
	\frac{\pi^2\sigma_0^2\kappa^{-1}R_0}{16D}\frac{1}{(\kappa R_0)^2}\times
\nonumber \\
&&	~~~\left\{\left[2-2(kR_0)^2\right]+
	\left[3+2(kR_0)^4-(kR_0)^2\right]\right\}
\eea
The first square bracket term simply adds a
small correction, $\alpha_{e}/4(\kappa R_0)^2$, to
the electrostatic surface tension energy.
Comparing the second square bracket term
with the bending energy change,
Eq. (\ref{Ebend}), one immediately
identifies this term as the ``electrostatic"
bending energy change  
with the corresponding ``electrostatic"
bending rigidity given by 
\beq
\gamma_{\rm e}=\pi\sigma_0^2\kappa^{-3}/4D.
\eeq
It is positive and, within a constant numerical factor,
agrees with the well-known expression\cite{ElBendingRigidity}
for $\gamma_{\rm e}$ calculated using 
other
methods. Thus expanding the electrostatic energy
with respect to the tube curvature ($1/\kappa R_0$)
in this near flat limit, one recovers all standard
formulae for the ``electrostatic" contributions
to the elastic parameters 
(surface tension and bending rigidity) 
of the bilayer surface.

Also in this limit, the relaxation energy
becomes, to the lowest order in $1/\kappa R_0$:
\beq
\Delta E_e^{\rm rel}\simeq-\varepsilon^2 \frac{\pi\alpha_e R_0}{16} 
	\frac{(kR_0)^4}{(\kappa R_0)^4}
\label{ErelaxStrong}
\eeq
%

Thus, the relaxation of charged molecules belong to the
fourth order or higher in the expansion
with respect to the tube curvature.
Since the electrostatic surface tension and the electrostatic
bending energy
are, correspondingly, the zeroth and second order terms
in this expansion
(the first order term in the expansion
vanishes because of the symmetry of the
reference flat surface),
charge relaxation energy gain is parametrically 
small compared to the electrostatic surface tension and 
bending energy in this limit and can be ignored. In other words,
in this strong screening limit, the charge density
of the surfactant bilayer can be considered
uniform during the deformation of the tube.

%

In the opposite limit of very weak screening,
$\kappa R_0 \ll 1$, generally speaking, the electrostatic
interaction is long range and is so large that
linear analysis becomes invalid in a very short time
after the instability develops and nonlinear terms
must be included in describing the development of
instability. This is, however, a very complicated task.
This is why in literature one
usually assumes the final (pearl-on-a-string) structure
of instability is given and variationally
minimized its total energy to find its parameters 
(size, period). Nevertheless, for the discussion
of the role of charge relaxation energy,
one can still use the result a linear analysis
given by exact expression, Eq. (\ref{x}), which is
valid at a very early time of instability.

Expanding the energies in powers of $\kappa R_0$,
for $k\geq \kappa$, we get to
the lowest order in $\kappa R_0$,
\bea
&&\Delta E_e^{\rm rel} =\varepsilon^2
	\frac{2\pi^2\sigma_0^2R_0^2}{D}
	\frac{I_0(kR_0)
	\left[K_0(kR_0)-kR_0K_1(kR_0)\right]^2}{K_0(kR_0)}
	\nonumber \\
&&\Delta E_e =\varepsilon^2
	\frac{2\pi^2\sigma_0^2R_0^2}{D}
	\left\{\frac{}{}
	\frac{(kR_0)^2}{2}
	\left[\frac{}{}I_0(kR_0)K_0(kR_0)-
	\right.\right.
\nonumber\\&&~~~
	\left.
	I_1(kR_0)K_1(kR_0)-\frac{1}{2}\right]+
	\frac{I_0(kR_0)}{K_0(kR_0)}
	kR_0K_1(kR_0)
	\times
\nonumber\\&&~~~
	\left.
	\left[K_0(kR_0)-kR_0K_1(kR_0)\frac{}{}\right]
	\right\}.
\label{kgreater}
\eea
Since for $k \geq \kappa$, all length scales are smaller
than the screening radius and electrostatic
interactions are not screened. Correspondingly, the energies are
independent of $\kappa$ as shown by the above 
equation (\ref{kgreater}). 

For $k\ll \kappa$, we get to the second lowest order in $\kappa R_0$
and $k/\kappa$:
\bea
\Delta E_e^{\rm rel} &=&\varepsilon^2
        \frac{2\pi^2\sigma_0^2R_0^2}{D}\left\{
	\ln\frac{\kappa R_0}{2}+(2+\gamma_E)+
\right.
\nonumber\\
&&~~~~~\left.
	\frac{1}{2}\frac{k^2}{\kappa^2}-
	\frac{k^2}{\kappa^2}
	\frac{\gamma_E}{\ln(\kappa R_0/2)}
	\right\}
	\nonumber \\
\Delta E_e &=&\varepsilon^2
        \frac{2\pi^2\sigma_0^2R_0^2}{D}
	\left[1-\frac{k^2}{\kappa^2}
	\frac{\gamma_E}{\ln(\kappa R_0/2)}\right].
\label{ksmaller}
\eea
Here $\gamma_E=0.5772$ is the Euler's constant.
For the later case, we see that the relaxation energy 
is larger
than the total electrostatic energy by a large logarithmic
term, $\ln(\kappa R_0/2)$. This is because,
the first expansion term which logarithmically 
diverges with $\kappa R_0\rightarrow 0$ in the
relaxation energy is exactly
equal in magnitude and opposite in sign to the
the first expansion term for the non-relaxation
energy. As a result, the total electrostatic
energy, contains only the
second and higher order expansion terms. Thus, it
is parametrically smaller than either of these components.
Obviously, within linear analysis,
the inclusion of the relaxation energy
is important in this limit to get the correct behaviour
of the electrostatic energy.



For the experimental situation of Ref. \onlinecite{Ajay}
where the screening radius, the tube radius and 
the pearl size are comparable to each other,
the situation is marginal.
Therefore, one might ask whether or not the charge relaxation
still plays a significant role.
To answer this question, we numerically evaluate
the exact (within linear analysis) expressions, 
Eqs. (\ref{EelecNorelax}) and (\ref{EelecRelax}),
for the $\Delta E_e^{\rm norel}$ and
$\Delta E_e^{\rm rel}$, respectively.
\begin{figure}[h]
\resizebox{8cm}{!}{\includegraphics{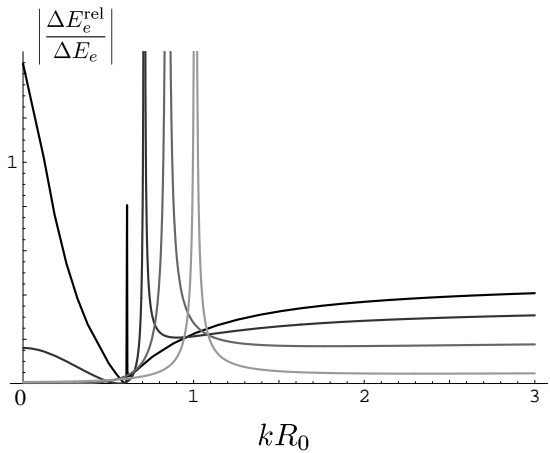}}
\caption{The absolute value of the 
ratio between the electrostatic
energy gain due to the relaxation of charge molecules
to the total electrostatic energy when a tube
deforms into a pearling structure as a function
of the wavevector of deformation, $kR_0$ for
$\kappa R_0=1$. Four different values of $\kappa R_0=0.1$,
 0.5, 1.0 and 2.0, are used. Lighter curve corresponds to
 higher $\kappa R_0$. The divergencies observed near $kR_0\sim 1$ is
due to the vanishing of $\Delta E_e$ owing to Raleigh instability.}
\label{fig:relaxratio}
\end{figure}
In Fig. \ref{fig:relaxratio},
we plot the ratio between the energy gained
due to the relaxation of charged molecules
and the total electrostatic energy change
for different values of $\kappa R_0$.
The divergence of this ratio observed at
about $kR_0\simeq 1$ is due to fact that the
total electrostatic energy change goes through
zero at this wavevector. As one can see from this
figure, for strong screening $\kappa R_0 > 1$,
the relaxation energy contributes a small part
of the total electrostatic energy while for weak
screening $\kappa R_0 < 1$, it contributes
significantly to the total energy. However, for the
marginal case $\kappa R_0=1$ which is more
relevant to the experiments,
the numerical result shows that
%
the relaxation of charged molecules contributes
about 20\%, which is a small fraction.
For smaller screening radius, the exponent of 4 in
Eq. (\ref{ErelaxStrong}) quickly reduces the relaxation energy
to an irrelevant contribution in the total energy.
Thus, one concludes that in the experimental situation
of Ref. \onlinecite{Ajay}, the electrostatic
relaxation is a small effect and the pearling
of the tubes is dominantly due to the Raleigh instability.


Before concluding, let us come back to the assumption made
in the previous section that the surface charge is always
in equilibrium with a given shape of the tube. To show that
this is a reasonable assumption, let us estimate
the charge relaxation of the surfactant bilayer.
Indeed, one can view the tube
as an RC circuit. One typical wavelength $k^{-1}\simeq R_0$,
the conductance of the circuit is the same as the conductivity,
${\cal R}^{-1}\simeq \sigma_0 e\mu$ with
$\mu\simeq 10^9m/sN$ the mobility of the surfactant.
Because all charges are screened at the distance $\kappa^{-1}$,
the capacitance of this circuit is of the order
${\cal C}\simeq R^2\kappa$. Thus the relaxation time
of this circuit (or of our surfactant tube) is of the order
$({\cal RC})^{-1}\simeq \sigma_0e\mu/R^2\kappa$ which
is about 0.1ms using experimental parameters. This is much
smaller than the growth rate of the pearling instability
(in seconds). Thus the surfactant charges, to a good approximation,
can be considered always in equilibrium.

\section{Conclusion}
We have considered the relaxation of charged surfactant in
the pearling instability for surfactant tubes observed
in the experiments of Ref. \onlinecite{Ajay}.
This effect is known to be the main driving force
for pearling instability of polyelectrolytes in 
poor solvent for which screening is very weak.
Using linear analysis, we showed that
for the marginal situation of Ref. \onlinecite{Ajay},
the effect is small and the pearling instability
is mainly due the Raleigh instability
caused by the surface tension of the bilayer.

Charge relaxation becomes important only for weak screening
such that $\kappa R_0 \ll 1$. In this limit, in our
linear analysis, the relaxation
energy is larger than the total electrostatic energy
by a logarithmic factor, $\ln(\kappa R_0/2)$. For
very weak screening condition, its inclusion is a must if 
one wants to obtain the correct
behaviour of electrostatic energy. 
Thus, although charge relaxation plays small role in
the biological phenomena of Ref. \onlinecite{Ajay},
it might well play a role in the domain of
flowing microemulsions, where the tube diameters
are much smaller.

\begin{acknowledgments}

The authors would like to thank A. Gopinathan, B. I. Shklovskii,
W. Zhang, S. Nagel, L. Silbert and L. Kadanoff for useful discussions.
This work was supported by the University of Chicago MRSEC program 
of the NSF (DMR-0213745). 
The experimental apparatus was made possible
by a NSF CRIF/Junior Faculty Grant (CHE-9816513). A.G. was partially
supported by the US-Israel Binational Foundation (2002-271). 
K.Y.C.L. is grateful for the support from the March of Dimes 
(\#6-FY03-58)
 and the Packard Foundation (99-1465). 

\end{acknowledgments}


\begin{thebibliography}{99}
\bibitem{ribbon}It should be mentioned here that
the observed tube structures can actually be ribbons of bilayer
surfactant. However, there are no known examples of
a ribbon structure undergoing pearling instability. 
This paper deals with pearling structures only.
For these structures, we believe the original structures are
cylindrical tubes.
\end{thebibliography}
\end{document}